\documentstyle[12pt,equations]{article}
\newcommand{\be}{\begin{equation}}
\newcommand{\ee}{\end{equation}}
\newcommand{\bea}{\begin{eqnarray}}
\newcommand{\eea}{\end{eqnarray}}
\newcommand{\bel}{\begin{eqalignno}}
\newcommand{\eel}{\end{eqalignno}}
\newcommand{\bes}{\begin{subequations}}
\newcommand{\ees}{\end{subequations}}

\newcommand{\prd}[3]{{Phys. Rev. D#1 (19#2) #3}}
\newcommand{\prl}[3]{{Phys. Rev. Lett. #1 (19#2) #3}}
\newcommand{\pla}[3]{{Phys. Lett. A#1 (19#2) #3}}
\newcommand{\plb}[3]{{Phys. Lett. B#1 (19#2) #3}}
\newcommand{\npb}[3]{{Nucl. Phys. B#1 (19#2) #3}}

\def\A{{\bf A}}
\def\B{{\bf B}}
\def\E{{\bf E}}
\newcommand{\Bo}{{\bf B}_0}
\newcommand{\Eo}{{\bf E}_0}
\newcommand{\C}{{\cal E\hspace{-8pt}E\hspace{-8pt}E}_0}

\def\a{{\bf a}}
\newcommand{\vb}{{\bf v}}
\newcommand{\vo}{{\bf v}_0}
\newcommand{\r}{{\bf r}}

\newcommand{\y}{{\bf y}}
\newcommand{\R}{{\bf R}}

\newcommand{\drdt}{\frac{d^2{\bf r}^*}{dt^{*2}}}

\newcommand{\prt}{\phi^{a}({\bf r},t)}
\newcommand{\airt}[1]{A^{a}_#1({\bf r},t)}
\newcommand{\aart}{A^{a}_0({\bf r},t)}

\newcommand{\ps}{\phi^{a}({\bf r}^*,t^*)}
\newcommand{\ais}[1]{A^{a}_#1({\bf r}^*,t^*)}
\newcommand{\as}{A^{a}_0({\bf r}^*,t^*)}

\newcommand{\pas}{\phi^{*a}({\bf r},t)}
\newcommand{\aias}[1]{A^{*a}_#1({\bf r},t)}
\newcommand{\aas}{A^{*a}_0({\bf r},t)}

\newcommand{\ptil}{\tilde{\phi}^{a}({\bf r}';\beta)}
\newcommand{\aitil}[1]{\tilde{A}^{a}_#1({\bf r}';\beta)}
\newcommand{\atil}{\tilde{A}^{a}_0({\bf r}';\beta)}

\newcommand{\pbo}{\bar{\phi}^{a}({\bf r};\beta)}
\newcommand{\aibo}[1]{\bar{A}^{a}_#1({\bf r};\beta)}
\newcommand{\abo}{\bar{A}^{a}_0({\bf r};\beta)}

\newcommand{\pbb}{\bar{\phi}^{a}({\bf r}';\beta)}
\newcommand{\aib}[1]{\bar{A}^{a}_#1({\bf r}';\beta)}
\newcommand{\ab}{\bar{A}^{a}_0({\bf r}';\beta)}

\newcommand{\vv}{\frac{1}{\sqrt{1-{{\bf v}_0}^2}}}
\newcommand{\ov}{\left(1-{{\bf v}_0}^2\right)}

\def\eijk{\epsilon^{ijk}}

\def\Rh{\hat{\R}}
\def\rhp{\hat{\r}'}
\def\rhpk{\hat{\r}'_k}
\def\rhpi{\hat{\r}'_i}
\def\rhpj{\hat{\r}'_j}
\def\rhpa{\hat{\r}'_a}
\def\vret{\vb_{ret}}

\begin{document}
\begin{flushright}SNUTP 93-97, hep-th/9402049 \end{flushright}

\vskip 1cm

\begin{center}
{\Large\bf BPS~Dyon~in~a~Weak~Electromagnetic~Field:\\
Equations of Motion and Radiation Fields}

\vskip 1cm
Dongsu Bak

{\it Center for Theoretical Physics, Laboratory for Nuclear Science and\\
Department of Physics, Massachusetts Institute of Technology,\\
Cambridge, MA02139, USA.}\\[5mm]

Choonkyu Lee

{\it Department of Physics and Center for Theoretical Physics,\\
Seoul National University, Seoul, 151-742, Korea}
\end{center}

%
%
\vspace{5mm}
\begin{center}
{\large\bf Abstract}\\[3mm]
\end{center}
\indent\indent
Dynamics of a BPS dyon in a weak, constant, electromagnetic field is
studied through a perturbative analysis of appropriate non-linear field
equations. The full Lorentz force law for a BPS dyon is established.
Also derived are the radiation fields accompanying the motion.

\newpage

The $SU(2)$ Yang-Mills-Higgs theory, spontaneously broken to $U(1)$ by an
adjoint Higgs field, allows non-singular time-independent classical solutions
of finite energy that carry nonzero magnetic charge \cite{1,2}. The same theory
also admits dyons, solutions that carry both electric and magnetic charges
\cite{3}.
In the so-called Prasad-Sommerfield limit \cite{4} of vanishing scalar field
potential, these monopoles or dyons become solutions of the first-order
Bogomolny equations \cite{5} or their suitable generalization \cite{6}.
A large number of papers have been written so far on these
Bogomolny-Prasad-Sommerfield(BPS) monopoles, but most of them considered
the {\em static} properties of the solutions only. A major exception here
is the work by Manton and others \cite{7}, in which the scattering of slowly
moving
BPS monopoles is studied with the help of a geodesic approximation in the
moduli
space of static multi-monopole solutions. Very recently \cite{8}, we have
also studied another important time-dependent phenomenon related to an
accelerating BPS monopole, that of radiation made up of massless fields in the
system. This involved a direct perturbative analysis of nonlinear field
equations(also initiated by Manton \cite{9}), and in the present work we report
some additional results obtained by the same method.

The behavior of an electrically-neutral BPS monopole in a weak magnetic field
${\bf B}_0$ is known already from ref. \cite{8}. We will here consider more
general problem, i.e., the behavior of a BPS dyon (with magnetic charge $g$ and
electric charge $q$ ) in the presence of a weak, asymptotically uniform,
electric and magnetic fields (${\bf E}_0$, ${\bf B}_0$). Following ref.
\cite{8},
we will then have to construct an appropriate time-dependent solution to the
nonlinear field equations, which describes the acceleration of the dyon and
accompanying field deformations and radiation simultaneously. (For a more
phenomenological approach to the related problem, see ref. \cite{10}).
To the first order in acceleration, we shall below set up a judicious
perturbative scheme and then solve the necessary differential equations
explicitly to exhibit the above-mentioned phenomena. A byproduct of our
analysis is that the nonlinear field equations of the Yang-Mills-Higgs theory
indeed require the center of a BPS dyon to move according to the naively
expected equation of motion
\be
\frac{d}{dt}\left(\frac{M_D {\bf v}}{\sqrt{1-\vb^2}} \right)
=g({\bf B}_0 -\vb\times\Eo)+q(\Eo + \vb\times\Bo)
\ee
Note that our analysis is entirely in the classical field theory context,
and so the electric charge $q$ is not quantized. We set $c=1$, and our metric
convention is that with signature$(-+++)$.

We shall specify our system first. The Lagrangian density is $(a=1,2,3)$
\be
{\cal L}=-\frac{1}{4}G_a^{\mu\nu}G_{\mu\nu a} -
 \frac{1}{2}(D^\mu \phi)_a(D_\mu \phi)_a\ ,
\ee
where
\bel
G_a^{\mu\nu}&=\partial^\mu A^\nu_a -\partial^\nu A^\mu_a +
e \epsilon_{abc}A^\mu_b A^\nu_c\ , \nonumber\\
(D_\mu\phi)^a&=\partial_\mu\phi^a + e \epsilon_{abc} A_{b\mu}\phi_c
               \equiv D_\mu^{ac}\phi_c\,,
\eel
and the Higgs fields are subject to the asymptotic boundary condition
$\phi^a \phi^a \rightarrow f^2 (\neq 0)$ as $r\rightarrow\infty$.
The field equations read
\bes
\bel
D_\mu^{ac}G_c^{\mu\nu}&=-e\epsilon_{abc}(D^\nu\phi)^b\phi^c\ ,\\
(D^\mu D_\mu \phi)^a &=0\ .
\eel
\ees
If we here define the electric and magnetic charges by
\be
q=\oint_{r=\infty} dS_i \frac{\phi^a}{|\mbox{\boldmath $\phi$}|}E_i^a\ ,\quad
g=\oint_{r=\infty} dS_i \frac{\phi^a}{|\mbox{\boldmath $\phi$}|}B_i^a\
\ee
(with $E_i^a\equiv G^{0i}_a$, $B_i^a\equiv \frac{1}{2}\epsilon_{ijk}G_a^{jk}$),
$g$ must be of the form $\frac{4\pi}{e}n$ for some integer $n$ by a
topological reason while $q$ can take on any continuous value.
Remarkably, for given $g$ and $q=g\tan\beta$, the static solutions to
eqs.(4a) and (4b) with the lowest possible energy can be obtained by
solving instead the first-order equations \cite{5,6}
\be
B_i^a=\mp \cos\beta\ (D_i\phi)^a\ ,\quad E_i^a=\mp\sin\beta\ (D_i\phi)^a \
,\quad
(D_0\phi)^a = 0\ .
\ee
These are equations describing static BPS dyons and for $\beta=0$
reduce to the Bogomolny equations for uncharged BPS monopoles:
\be
B_i^a = \mp (D_i\phi)^a\ ,\quad A_0^a =0\ .
\ee

Actually all dyon solutions to eq.(6), denoted as $(\bar{\phi}^a(\r;\beta),\
\bar{A}_i^a(\r;\beta),\ \bar{A}_0^a(\r;\beta))$, can be obtained from the
static monopole solutions $(\bar{\phi}^a(\r;\beta=0),\
\bar{A}^a_i(\r;\beta=0))$
satisfying  eq.(7). This is achieved by the simple substitution \cite{11}
\bes
\bel
\bar{\phi}^a(\r;\beta)
  &=\left[\bar{\phi}^a(\y;\beta=0)\right]_{\y=\r\cos\beta}\ , \\
\bar{A}^a_i(\r;\beta)
  &=\cos\beta\ \left[\bar{A}^a_i(\y;\beta=0)\right]_{\y=\r\cos\beta}\ ,\\
\bar{A}^a_0(\r;\beta)
  &=\mp\sin\beta\ \left[\bar{\phi}^a(\y;\beta=0)\right]_{\y=\r\cos\beta}
\nonumber\\
&=\mp\sin\beta\ \bar{\phi}^a(\r;\beta)\ .
\eel
\ees
Especially, by making this substitution with the well-known BPS one-monopole
solution
(with $g=\mp\frac{4\pi}{e}$ and mass $M=\frac{4\pi}{e}f$)
\bel
\bar{\phi}^a(\r;\beta=0)&=\pm\hat\r_af
\left[\coth m_v r -\frac{1}{m_v r}\right]\ ,\quad (m_v=ef) \nonumber\\
\bar{A}^a_i(\r;\beta=0)&= \epsilon_{ail}\frac{\hat\r_l}{er}
\left[1 - \frac{m_v r}{\sinh m_v r}\right]\ ,
\eel
we immediately obtain the spherically symmetric BPS dyon solution \cite{4}
with $g=\mp\frac{4\pi}{e}\, , q=\mp\frac{4\pi}{e}\tan\beta $ and mass
$M_D = \frac{4\pi}{e}\frac{f}{\cos\beta}$. In this paper we will focus
on time-dependent dynamics of this dyon when there is a weak uniform
electromagnetic
field asymptotically, viz., under the condition that
\be
r\rightarrow \infty :
 \quad\frac{\phi^a}{|{\bf \phi}|}B_i^a \rightarrow (\Bo)_i\ ,\quad
\frac{\phi^a}{|{\bf \phi}|}E_i^a \rightarrow (\Eo)_i\ ,
\ee
It is impractical to look for exact solutions to the full field equations (4a)
and (4b)
appropriate to this circumstance. But, to the first order in $\Bo$ and $\Eo$
(and hence
to the first order in the dyon acceleration ${\bf a}$ also), the corresponding
approximate
solutions to eqs.(4a) and (4b) can be constructed explicitly.

Let us first generalize the result of ref.~\cite{8} for a BPS monopole (with
$q=0$) by
allowing the asymptotic electric field $\Eo$ as well. Assuming that the
monopole has zero
velocity at $t=0$, we may then write the ansatz
\bel
\phi^a(\r,t)&=\bar{\phi}^a(\r';\beta=0) + \Pi^a(\r')\hspace{3mm}
 (\equiv\tilde{\phi}^a(\r';\beta=0))\ , \nonumber\\
\A^a_i(\r,t)&=\bar{A}^a_i(\r';\beta=0) + \alpha_i^a(\r')\hspace{3mm}
(\equiv\tilde{A}^a_i(\r';\beta=0))\ ,\nonumber\\
\A^a_0(\r,t)&=-t a_i\bar{A}^a_i(\r';\beta=0) + \alpha_0^a(\r')
\eel
with $\r'\equiv \r-\frac{1}{2}\a t^2$\ , $\a$ being the expected acceleration
of the monopole. Here, $(\bar{\phi}^a,\bar{A}^a_i)$ denotes the BPS
one-monopole solutions in eq.(9) and the presence of the term
$\alpha_0^a(\r')$ distinguishes the present form from that of ref.~\cite{8}.
Our approximation is to neglect any terms
beyond $O(a)$ (here $a\equiv|\Bo|,\ |\Eo|$ or $|\a|$) in $\Pi^a$,
$\alpha_i^a$ and $\alpha_0^a$, and in this approximation the consistency of
our assumption that these functions depend on time only through $\r'$ can be
established. Then, from the
field equations (4a) and (4b), we can deduce the following set of equations:
\bes
\bel
&(\bar{D}^i \alpha^j)^a - (\bar{D}^j \alpha^i)^a = \mp\epsilon^{ijk}
\left(\bar{D}_k^{ab}\Pi^b + e\epsilon_{abc}\alpha^b_k\bar{\phi}^c +
a_k\bar{\phi}^a \right) \ ,\\
&(\bar{D}_i\bar{D}_i\alpha_0)^a =
-e^2\epsilon_{abc}\epsilon_{bdf}\bar{\phi}_c\bar{\phi}_f\alpha_0^d\ ,
\eel
\ees
where we have defined
$\bar{D}_i^{ac}=\partial_i\delta_{ac}+e\epsilon_{abc}\bar{A}_i^b$. Since
$\bar B_i^a\equiv \frac{1}{2}\epsilon_{ijk}\bar G_a^{jk} =
\mp(\bar{D}_i\bar{\phi})^a$,
eq.(12a) is equivalent to \cite{9}
\be
B_i^a=\mp (D_i + a_i)^{ab}\phi_b\ ,
\ee
while the field strength $E_i^a$ can be expressed as
\be
E_i^a=-ta_j\bar{G}^{ij}_a + (\bar{D}_i\alpha_0)^a\ .
\ee
Then, comparing these with our asymptotic requirement (10), we can immediately
conclude
that
\begin{enumerate}
\item[(i)] $\Bo=\mp\a f$ or, equivalently, $M\a=g\Bo$ ($M=\frac{4\pi}{e}f$
is the monopole mass), and
\item[(ii)] $\alpha_0^a(\r')\rightarrow \pm\Eo\cdot\r'\hat\r'_a$
as $r'=|\r'|\rightarrow\infty$.
\end{enumerate}

The functions $\Pi^a(\r')$ and $\alpha_i^a(\r')$, which are solutions to
eq.(12a),
have been found already in ref.~\cite{8}. Using those results, we can equate
$\phi^a(\r,t)$ and $A_i^a(\r,t)$ to the first order in $\Bo$ and $\Eo$ to the
expressions
(here $\a=\mp\Bo/f$)
\bes
\bel
\tilde{\phi}^a(\r';\beta=0)&=\pm\hat\r_a'
\left[f\left(\coth m_v r' -\frac{1}{m_v r'}\right) +
\frac{1}{2e}\hat\r'\cdot\a\left(1-\frac{m_v r'}{\sinh m_v r'} \right)
 \right]\pm \frac{1}{2} a_a\frac{fr'}{\sinh m_v r'}\,,\nonumber\\& \\
\tilde{A}^a_i(\r';\beta=0)&= \epsilon_{aij}\frac{\hat\r_j'}{er'}
\left(1 - \frac{m_v r'}{\sinh m_v r'}\right)+
\frac{fr'}{2}\left(\coth m_v r' -\frac{1}{m_v r'}\right)
\epsilon_{ijk}\hat\r_j'a_k\hat\r_a'\,,
\eel
\ees
respectively. Remarkably the closed-form solution to eq.(12b) which satisfies
the
above asymptotic boundary condition can also be found:
\be
\alpha_0^a(\r')=\pm(\coth m_v r')\Eo\cdot\r'\hat\r_a'
\pm \frac{r'}{\sinh m_v r'}\left[(\Eo)_a-(\Eo\cdot\hat{\r}')\hat\r'_a\right]\ .
\ee
The functions $\alpha_0^a(\r')$ here describe necessary field deformations in
the
presence of an asymptotic electric field $\Eo$; but, in association with this
term, there is no additional electromagnetic or Higgs scalar radiation (beyond
those already present in association with the asymptotic magnetic field $\Bo$).
On the other hand, this solution in the presence of nonzero $\Bo$ and $\Eo$
will be
important to discuss the Lorentz boost property. See below.

If one considers a Lorentz boost transformation of the solution specified by
eqs.(11), (15a), (15b) and (16), viz.,
\bel
&\pas=\ps\ ,\nonumber\\
&\aias{i}=\ais{i}+\left(\vv-1\right)(\hat\vb_0^j\ais{j})\hat\vb_0^i
  -\vv v_0^i\as\ , \nonumber\\
&\aas=\vv\left\{\as-v_0^j\ais{j}\right\}\ ,\\
&\left(\r^*=\r+\left(\vv-1\right)(\hat\vb_0\cdot\r)\hat\vb_0-\vv\vo t\ ;
   \hspace{3mm}
t^*=\frac{t-\vo\cdot\r}{\sqrt{1-\vo^2}} \right)\nonumber
\eel
the result will be another solution to the field equations. It describes
an accelerating BPS monopole with finite initial velocity $\vo$ in the
presence of a uniform electromagnetic field $(\Eo^*,\Bo^*)$, where
\bel
\Eo^*&=(\hat\vb_0\cdot\Eo)\hat\vb_0
+\vv\left[\Eo-(\hat\vb_0\cdot\Eo)\hat\vb_0-\vo\times\Bo\right]\ ,\nonumber\\
\Bo^*&=(\hat\vb_0\cdot\Bo)\hat\vb_0
+\vv\left[\Bo-(\hat\vb_0\cdot\Bo)\hat\vb_0+\vo\times\Eo\right]\ .
\eel
[Especially, by choosing $\Bo=\vo\times\Eo$, one can obtain the solution
appropriate to a BPS monopole with a finite initial velocity $\vo$ in the
presence of some nonzero electric field $\Eo^*$ (but $\Bo^*=0$).]
The trajectory assumed by the center of this monopole will be determined
by the condition $\r^*-\frac{1}{2}\left(\frac{g\Bo}{M}\right)(t^*)^2=0$,
or by
\bel
\drdt &= \frac{g}{M}\Bo \nonumber\\
       &= (\hat\vb_0\cdot\B^*)\hat\vb_0
+\vv\left[\Bo^*-(\hat\vb_0\cdot\Bo^*)\hat\vb_0-\vo\times\Eo^*\right]\ ,
\eel
where we have used eq.(18). At the same time, we have
\be
\drdt= \frac{(\hat\vb_0\cdot\frac{d\vb}{dt})\hat\vb_0}{\ov^{3/2}}
       + \frac{1}{\ov}\left\{\frac{d\vb}{dt}
       -\left(\hat\vb_0\cdot\frac{d\vb}{dt}\right)\hat\vb_0
\right\} + O(a^2) \ ,
\ee
(here, $\vb(t)=\frac{d\r(t)}{dt}$ ($=\vo + O(a)$) denotes the velocity of the
monopole),
and consequently eq.(19) can be recast as
\be
\frac{d}{dt}\left(\frac{M\vb}{\sqrt{1-\vb^2}}\right)
= g\left[\Bo^*-\vb\times\E^*\right] + O(a^2)\ .
\ee
Eq.(1) is thus confirmed for an uncharged monopole.

A similar analysis can also be carried out for a BPS dyon, at least to the
first
order in the asymptotic electromagnetic fields $\Eo$ and $\Bo$ (and hence in
the
dyon acceleration $\a$). Again choosing the reference frame in such a way that
the dyon may have zero velocity at $t=0$, we may generalize the ansatz (11) as
\bel
\prt&=\ptil\ ,\nonumber\\
\airt{i}&=-ta_i\ab + \aitil{i}\ ,\\
\aart&=-ta_i\aib{i}+\atil\nonumber
\eel
with
\bel
\ptil&=\pbb +\Pi^a(\r';\beta)\ ,\nonumber\\
\aitil{i}&=\aib{i} + \alpha_i^a(\r';\beta)\ ,\\
\atil&=\mp\sin\beta\ \ptil+ \alpha^a_0(\r';\beta)\ ,\nonumber
\eel
where $\r'=\r-\frac{1}{2}\a t^2$. Here the functions $\pbo$, $\aibo{i}$ and
$\abo$ represent the static dyon solution in eqs.(8a)-(8c), and the
yet-to-be-determined functions $(\Pi^a, \alpha_i^a, \alpha_0^a)$ are assumed to
be $O(a)$. Notice the specific way the time coordinate $t$ enters our ansatz.
The appearance of the piece $-ta_i\ab$ or $-ta_i\aib{i}$ in eq.(22) may well be
understood from the viewpoint of the instantaneous Lorentz boost. Also
see the relationship (8c) for a possible motivation for writing $\atil$ as
in eq.(23).

The consistency of the above ansatz can be verified explicitly; inserting the
ansatz (22) into the full field equations leads to differential equations
whose sole dependent variables are $\r'$. Explicitly, we obtain from eq.(4a)
\bes
\bel
&(\tilde D_j+a_j)^{ac}\tilde G_c^{ji}
 +e\epsilon_{abc}\tilde A_0^b(\tilde D_i\tilde A_0)^c
 =-e\epsilon_{abc}(\tilde D_i\tilde\phi)^b\tilde\phi^c\,,\\
&\tilde D_i^{ab}(\tilde D_i\tilde A_0+a_i\tilde A_0)^b
 +e^2\epsilon_{abc}\epsilon_{bdf}\tilde\phi^c\tilde A_0^d\tilde\phi^f=0\,,
\eel
and from eq.(4b)
\be
\tilde D_i^{ab}(\tilde D_i\tilde\phi+a_i\tilde\phi)^b
 +e^2\epsilon_{abc}\epsilon_{bdf}\tilde A_0^c\tilde A_0^d\tilde\phi^f=0\,,
\ee
\ees
where $\tilde D_i^{ab}=[D_i^{ab}]_{A_i^a\rightarrow\tilde A_i^a}$,
$\tilde G_c^{ji}=[G_c^{ji}]_{A_i^a\rightarrow\tilde A_i^a}$,
and the suppressed dependent variable is $\r'$.  Then, identifying $\atil$
as in eq.(23), we found that these three equations are satisfied (to $O(a)$)
as long as the following conditions are fulfilled:
\bes
\bel
&\tilde B_i^a=\mp(\tilde D_i+a_i)^{ab}(\cos\beta\ \tilde\phi^b
 \pm\tan\beta\ \alpha_0^b)\,,\\
&(\bar D_i\bar D_i\alpha_0)^a
 =-e^2\cos^2\beta\
\epsilon_{abc}\epsilon_{bdf}\bar\phi^c\bar\phi^f\alpha_0^d\,.
\eel
\ees
Notice that, for $\beta = 0$, these reduce to the corresponding equations for
an
uncharged BPS monopole, i.e., eqs.(12b) and (13).
In the present case the field strength $E_i^a$ to $O(a)$ becomes
\be
E_i^a(\r,t)=-ta_j\bar G_a^{ij}+(\tilde D_i+a_i)^{ab}\tilde A_0^b\,.
\ee

The acceleration $\a$ above is not arbitrary but fixed in terms of the given
asymptotic electromagnetic fields. For this purpose the asymptotic
condition (10) may be used with the field strengths given by eqs.(25a)
and (26).  Actually we will also assume that
$\frac{\phi^a}{|\mbox{\small\boldmath
$\phi$}|}(\tilde{D}_i\tilde{\phi})^a\rightarrow 0$ as
$r'\rightarrow\infty$; this can be viewed as our additional asymptotic
condition.
Moreover, according to our lesson form the uncharged monopole case, it is
quite natural to suppose that $\alpha_0^a(\r';\beta)\rightarrow\cos\beta\
\C\cdot\r'\hat\r_a'$ (for some constant vector $\C$) as
$r'\rightarrow\infty$. Now, from eq.(25a) and the asymptotic conditions,
\be
\Bo = \mp f\cos\beta\; \a \mp \sin\beta\;\C\ ,
\ee
while eq.(26) and the asymptotic conditions lead to
\be
\Eo=\mp f\sin\beta\;\a \pm \cos\beta\;\C\ .
\ee
We thus have
\be
\C = \mp(\sin\beta\;\Bo -\cos\beta\;\Eo)\ ,\qquad
\a = \mp\frac{1}{f}(\cos\beta\; \Bo + \sin\beta\;\Eo)\ ,
\ee
and especially, recalling that the dyon has mass
$M_D=\frac{4\pi}{e}\frac{f}{\cos\beta}$,
$g=\mp\frac{4\pi}{e}$ and $q=\mp\frac{4\pi}{e}\tan\beta$, the second of these
can be recast as
\be
 M_D\a=g\Bo+q\Eo\ .
\ee
This corresponds to the formula (1) in dyon's instantaneous rest frame.
The equation of motion in the general form (1) can be secured if the
Lorentz boost transformation (17) is considered for the above solution.
This proceeds in much the same way as in eqs.(17)-(21), and so we shall
not repeat it here.

We will now find the explicit solution to eqs.(25a) and (25b). Here it
is convenient to introduce rescaled quantities
\be
  \y =\r'\cos\beta\ ,\quad {\cal A}^a_i(\y)=\frac{1}{\cos\beta}\tilde{A}^a_i
\bigl(\r'=\frac{\y}{\cos\beta};\beta\bigr)\ .
\ee
Then, with the help of the relationships (8a)-(8c), we can rewrite
eqs.(25a) and (25b) as
\bes
\bel
&{\cal B}_i^a=\mp\left({\cal D}_i^{(y)}+\frac{a_i}{\cos\beta}\right)^{ab}
 \left(\tilde\phi^b\pm\frac{\sin\beta}{\cos^2\beta}\alpha_0^b\right)\,,\\
&(\bar D_i^{(y)}\bar D_i^{(y)}\alpha_0)^a
 =-e^2\epsilon_{abc}\epsilon_{bdf}\bar\phi^c(\y;\beta=0)
  \bar\phi^f(\y;\beta=0)\alpha_0^d\,,
\eel
\ees
where $\bar{D}_i^{(y)\,ac}\equiv\frac{\partial}{\partial y_i}\delta_{ac}
+e\epsilon_{abc}\bar{A}_i^b(\y;\beta=0)\ ,{\cal D}_i^{(y)\,ac}\equiv
\frac{\partial}{\partial y_i}\delta_{ac}+e\epsilon_{abc}{\cal A}^b_i(\y)$,
and ${\cal B}_i^a(\y)$ denotes the magnetic field strength obtained from
the vector potential ${\cal A}^a_i(\y)$. Eq.(32b) coincides with our
earlier equation(eq.(12b)), and so the solution with the desired
asymptotic behavior is immediately identified with (here $y=|\y| =r'\cos\beta$)
\be
\alpha_0^a=(\coth m_vy){\cal E\hspace{-8pt}E}_0\cdot\y\;\hat\y_a
 +\frac{y}{\sinh m_vy}[({\cal E\hspace{-8pt}E}_0)_a-({\cal
E\hspace{-8pt}E}_0\cdot\hat\y)\;\hat\y_a]\,.
\ee
Moreover, comparing eq.(32a) with eq.(13), we are led to write
\bel
{\cal A}_i^a &=[\tilde A_i^a(\y;\beta=0)]_{\a\rightarrow\frac{\a}{\cos\beta}}
              +\hat\alpha_i^a(\y)\,,\nonumber\\
\tilde\phi^a &=\mp\frac{\sin\beta}{\cos^2\beta}\alpha_0^a
        +[\tilde\phi^a(\y;\beta=0)]_{\a\rightarrow\frac{\a}{\cos\beta}}
        +\hat\Pi^a(\y)\,,
\eel
where $\hat{\alpha}_i^a(\y)$ and $\hat{\Pi}^a(\y)$ now satisfy the
homogeneous equation
(cf. eq.(12a))
\be
(\bar D^{(y)i}\hat\alpha^j)^a-(\bar D^{(y)j}\hat\alpha^i)^a
 =\mp\eijk(\bar D_k^{(y)ab}\hat\Pi^b
           +e\epsilon_{abc}\hat\alpha_k^b\bar\phi^c(\y;\beta=0))\,,
\ee
the functions $\tilde{\phi}^a(\y;\beta=0)$ and $\tilde{A}_i^a(\y;\beta=0)$ are
given explicitly in eqs(15a) and (15b), and $\alpha_0^a$ in eq.(33). Since the
function $\alpha^a_0$ increases linearly for large $r'$, we must here include
a suitable homogeneous solution $(\hat{\alpha}_i^a ,\hat{\Pi}^a)$ so that
our asymptotic condition $\displaystyle \lim_{r'\rightarrow\infty}
\frac{\phi^a}{|\mbox{\boldmath $\phi$}|}(\tilde D_i\tilde\phi)^a=0$
may be fulfilled.

The general, normalizable or nonnormalizable, solutions of eq. (35) have been
known for some time \cite{12}. Because of the above-mentioned asymptotic
condition, we may here demand $\hat\Pi^a(\y)$ to behave such as
\be
y\rightarrow\infty:\quad
 \hat\Pi^a(\y)
  \rightarrow\pm\frac{\sin\beta}{\cos^2\beta}{\cal
E\hspace{-8pt}E}_0\cdot\y\;\hat\y_a\,.
\ee
Then, using the result of ref. \cite{12}, we find that the appropriate
solution to eq. (35) is given by
\bel
\hat\alpha_i^a(\y)&=\epsilon_{ail}\frac{\partial}{\partial y^l}
  \left(\frac{y}{\sinh m_vy}V\right)
  +(1-\cosh m_vy)\hat\y_a\epsilon_{ilm}\hat\y_l\frac{\partial}{\partial y^m}
   \left(\frac{y}{\sinh m_vy}V\right)\,,\nonumber\\
\hat\Pi^a(\y)&=\mp\hat\y^a\frac{\partial}{\partial y}
   \left(y\coth m_vy V\right)\mp\left(\frac{\partial}{\partial y^a}
       -\hat\y_a\frac{\partial}{\partial y}\right)
  \left(\frac{y}{\sinh m_vy}V\right)
\eel
with $V=-\frac{\sin\beta}{2\cos^2\beta}{\cal E\hspace{-8pt}E}_0\cdot\y$. Based
on these
findings, the functions $\ptil$, $\aitil{i}$ and $\atil$ in eq. (22) are
explicitly given as
\bel
\ptil&=\pm\rhpa\left[f\left\{\coth[(m_v\cos\beta)r']
                   -\frac{1}{(m_v\cos\beta)r'}\right\}\right.\nonumber\\
   &\hspace{15mm}\left.+\frac{\rhp\cdot\a}{2e\cos\beta}\left\{
1-\frac{(m_v\cos\beta)r'}{\sinh[(m_v\cos\beta)r']}\right\}\right]\nonumber\\
 &\ \pm\frac{f}{2}a_a\frac{r'}{\sinh[(m_v\cos\beta)r']}
  \mp\frac12({\cal E\hspace{-8pt}E}_0\cdot\r')\rhpa
     \frac{(m_v\sin\beta)r'}{\sinh^2[(m_v\cos\beta)r']}\nonumber\\
 &\ \mp\frac{\tan\beta}2\left[({\cal E\hspace{-8pt}E}_0)_a-({\cal
E\hspace{-8pt}E}_0\cdot\rhp)\rhpa\right]
    \frac{r'}{\sinh[(m_v\cos\beta)r']}\;,\nonumber\\
\aitil{i}&=\epsilon_{aij}\frac{\rhpj}{er'}\left\{
     1-\frac{(m_v\cos\beta)r'}{\sinh[(m_v\cos\beta)r']}\right\}\nonumber\\
 &\ +\frac{f\cos\beta}{2}r'
      \left\{\coth[(m_v\cos\beta)r']-\frac1{(m_v\cos\beta)r'}\right\}
      \eijk\rhpj a_k\rhpa\nonumber\\
 &\ -\frac{\sin\beta}2\epsilon_{ail}\frac\partial{\partial {r'}^l}\left\{
     \frac{r'({\cal
E\hspace{-8pt}E}_0\cdot\r')}{\sinh[(m_v\cos\beta)r']}\right\}\nonumber\\
 &\ -\frac{\sin\beta}2
      \frac{r'\{1-\cosh[(m_v\cos\beta)r']\}}{\sinh[(m_v\cos\beta)r']}
      \rhpa\epsilon_{ilm}\hat\r'_l({\cal E\hspace{-8pt}E}_0)_m\,,\nonumber\\
\atil&=\mp\sin\beta\ \ptil
    +\cos\beta\coth[(m_v\cos\beta)r']{\cal
E\hspace{-8pt}E}_0\cdot\r'\hat\r'_a\nonumber\\
 &\ +\frac{(\cos\beta)r'}{\sinh[(m_v\cos\beta)r']}
     \left[({\cal E\hspace{-8pt}E}_0)_a-({\cal
E\hspace{-8pt}E}_0\cdot\rhp)\rhpa\right]
\eel
with the constant vectors ${\cal E\hspace{-8pt}E}_0$ and $\a$ determined by eq.
(29). Notice
that these functions are everywhere regular.

In eqs. (22) and (38) we have the explicit solution appropriate to an
accelerating BPS dyon. Its asymptotic behavior is of particular interest since
it carries information on emitted radiation. Let us concentrate on the
asymptotic behaviors of the corresponding gauge-covariant quantities. A short
calculation using eqs. (22) and (38), while ignoring exponentially small
terms, produces the following results (valid for $(m_v\cos\beta)r'\gg1$):
\bes
\bel
G_a^{ij}(\r,t)&\sim\eijk\left\{-\frac{1}{e{r'}^2}\rhpk
  +\frac{1}{2er'}[a_k+\rhpk(\rhp\cdot\a)]
  +\frac{\tan\beta}{e}\frac{t(\hat\r'\times\a)_k}{{r'}^2}\right.\nonumber\\
&\hspace{14mm}-f\cos\beta\ a_k-\sin\beta\ ({\cal
E\hspace{-8pt}E}_0)_k\Biggr\}\rhpa\,,\\
G_a^{0i}(\r,t)&\sim\left\{-\frac{\tan\beta}{e}\frac{\rhpi}{{r'}^2}
  +\frac{\tan\beta}{2e}[a_i+\rhpi(\rhp\cdot\a)]
  -\frac{t(\rhp\times\a)_i}{e{r'}^2}\right.\nonumber\\
&\hspace{8mm}-f\sin\beta\ a_i+\cos\beta\ ({\cal
E\hspace{-8pt}E}_0)_i\Biggr\}\rhpa\,,\\
(D^i\phi)^a(\r,t)&\sim\pm\frac{1}{e\cos\beta}\left\{
  \frac{\rhpi}{{r'}^2}+\frac{a_i-\rhpi(\rhp\cdot\a)}{2r'}\right\}\rhpa\,,\\
(D^0\phi)^a(\r,t)&\sim\pm\frac{1}{e\cos\beta}
  \frac{t\a\cdot\rhp}{{r'}^2}\rhpa\,.
\eel
\ees
The asymptotic electromagnetic fields also follow from these expressions by
setting $G_a^{ij}(\r,t)\sim\eijk B_k^{em}(\r,t)\rhpa$ and $G_a^{0i}(\r,t)\sim
E_i^{em}(\r,t)\rhpa$. Non-trivial expressions we have in eqs. (39c) and (39d)
can be interpreted as the long-range effects associated with the massless
Higgs scalar in this theory.

Now, as in Ref. \cite{8}, it will be convenient to rewrite the above
expressions using the ``retarded'' distance $\R=\r-\frac12\a t_{ret}^2$, where
$t_{ret}$ is determined (for given $\r$ and $t$) through the implicit equation
$t-t_{ret}=|\r-\frac12\a t_{ret}^2|\equiv R$. Here, noting that
$\frac12|\a|t_{ret}^2\ll R$ for sufficiently small acceleration, we have
\be
\r'=\R-\vb_{ret}R-\frac12\a R^2\,,\quad (\vb_{ret}\equiv\a t_{ret})
\ee
and $r'=R(1-\Rh\cdot\vb_{ret}-\frac12R\Rh\cdot\a)$. With these and the
identifications $g=\mp\frac{4\pi}{e}$ and $q=\mp\frac{4\pi}{e}\tan\beta$, it
is now a simple matter to recast the above asymptotic fields as
\bes
\bel
\B^{em}(\r,t)&\sim\frac{g}{4\pi}\frac{\Rh-\vret}{(1-\Rh\cdot\vret)^3R^2}
 +\frac{g}{4\pi}\frac{\Rh\times(\Rh\times\a)}{R}\nonumber\\
&\ \ -\frac{q}{4\pi}\frac{\Rh\times\vret}{R^2}
  -\frac{q}{4\pi}\frac{\Rh\times\a}{R}+\Bo\,,\\
\E^{em}(\r,t)&\sim\frac{q}{4\pi}\frac{\Rh-\vret}{(1-\Rh\cdot\vret)^3R^2}
 +\frac{q}{4\pi}\frac{\Rh\times(\Rh\times\a)}{R}\nonumber\\
&\ \ +\frac{g}{4\pi}\frac{\Rh\times\vret}{R^2}
  +\frac{g}{4\pi}\frac{\Rh\times\a}{R}+\Eo\,,\\
(D^i\phi)^a(\r,t)&\sim\left\{\frac{g_s}{4\pi}
   \frac{\Rh_i-v_{ret,i}}{(1-\Rh\cdot\vret)^3R^2}
  +\frac{g_s}{4\pi}\frac{(\Rh\cdot\a)\Rh_i}{R}\right\}\rhpa\,,\\
(D^0\phi)^a(\r,t)&\sim\left\{\frac{g_s}{4\pi}
 \frac{\Rh\cdot\vret}{R^2}+\frac{g_s}{4\pi}\frac{\Rh\cdot\a}{R}\right\}\rhpa\,,
\eel
\ees
where we have defined $g_s=\pm\frac{4\pi}{e\cos\beta}$, and $\Eo$ and $\Bo$
are the asymptotic uniform electromagnetic fields given through eqs. (27)
and (28).  In these expressions the $O(R^{-2})$ terms make the near-zone
fields associated with electromagnetic or Higgs scalar force.  Observe that
the near-zone fields we have obtained for $\B^{em}$ and $\E^{em}$ agree
with the naively expected results on the basis of the classical Maxwell
theory and the duality argument [13].  The same can be said also for the
radiation fields, which we discuss below.

The $O(R^{-1})$ terms in eqs.(41a)-(41d) describe the radiation fields
accompanying an accelerating BPS dyon. Explicitly we have
\bel
\B_{rad}=\frac{g}{4\pi}\frac{\Rh\times(\Rh\times\a)}{R}
        +\frac{q}{4\pi}\frac{\a\times\Rh}{R}\,,\nonumber\\
\E_{rad}=\frac{q}{4\pi}\frac{\Rh\times(\Rh\times\a)}{R}
        -\frac{g}{4\pi}\frac{\a\times\Rh}{R}
\eel
in connection with the electromagnetic interaction, and also the Higgs scalar
radiation described by
\be
(D^i\phi)_{rad}^a=\frac{g_s}{4\pi}\frac{(\Rh\cdot\a)\Rh_i}{R}\rhpa\,,\quad
(D^0\phi)_{rad}^a=\frac{g_s}{4\pi}\frac{\Rh\cdot\a}{R}\rhpa\,.
\ee
The duality is manifest in eq. (42); in fact, if one calculates the dyon
radiation fields by invoking the duality argument to the well-known radiation
fields of a uniformly accelerating electric charge, one ends up with eq. (42).
Also the expressions in eq. (43) (more precisely, the quantities at front of
$\rhpa$)  can be identified with the massless Klein-Gordon radiation fields
produced by a uniformly accelerating point scalar source with strength
$g_s=\pm\frac{4\pi}{e\cos\beta}$. Since the energy-momentum tensor is given by
\be
T_{\mu\nu}=(D^\mu\phi)_a(D^\nu\phi)_a+G_a^{\mu\lambda}G_{a\lambda}^\nu
           +\eta^{\mu\nu}{\cal L}\,,
\ee
the radiation energy flux can be found using
\bel
T_{rad}^{0i}&=[G_a^{0k}G_a^{ik}+(D^0\phi)^a(D^i\phi)^a]_{rad}\nonumber\\
  &=\E_{rad}^{em}\times\B_{rad}^{em}
   +(D^0\phi)_{rad}^a(D^i\phi)_{rad}^a\,.
\eel
Inserting eqs. (42) and (43) into this formula yields
\be
T_{rad}^{0i} = \frac{(g^2 + q^2)}{16 \pi^2 R^2} |\a \times \hat{\R}|^2
\hat\R_i + \frac{g_s^2}{16 \pi^2 R^2} |\a \cdot \hat{\R}|^2 \hat\R_i\,,
\ee
where the first term is the electromagnetic contribution, and the second
due to the Higgs scalar.

The dynamics of a BPS dyon in the presence of an asymptotically uniform
electromagnetic field has been successfully extracted using only field
equations of the Yang-Mills-Higgs system, at least in the small
acceleration regime. The given field equations are intrinsically nonlinear,
and they serve to specify the {\em entire} dynamics of a BPS dyon; the
situation here is quite analogous to the case of Einstein's field
equations [14].  The force law and radiation fields found by our analysis
support the duality between electric and magnetic quantities. But note that
quantum (or loop) effects have not been taken into account in our
discussion.

\begin{center}\bf Acknowledgements\end{center}
\ \indent
Part of this work was done when one of us(C.L.) was visiting Center for
Theoretical Physics, MIT during the summer of 1993, and he wishes to thank
Prof. R. Jackiw for the hospitality especially and the support by the NSF on
contract \#910653. The work of D.B. is
supported in part by funds provided by the U.S. Department of Energy
(D.O.E.) under contract \#DE-AC02-76ER03069.  The work of C.L. is supported
in part by the Korean Science and Engineering Foundation (through the
Center for Theoretical Physics, SNU) and the Ministry of Education,
Republic of Korea.

\end{document}